# How to Capture and Study Conversations Between Research Participants and ChatGPT: GPT for Researchers (g4r.org)

Jin Kim

D'Amore-McKim School of Business, Northeastern University

**Author Note**

Jin Kim  https://orcid.org/0000-0002-5013-3958

This paper and the tool (website) being introduced (g4r.org) were written with assistance from ChatGPT. Additional materials and possibly the code for G4R (e.g., PHP, JS, and HTML code that run g4r.org) will be openly available at the project's Open Science Framework page, https://osf.io/xn8hc/?view_only=c624157ae6af437db3014db694f3382a, prior to a peer-reviewed publication of this paper. We have no conflicts of interest to disclose.

Correspondence concerning this article should be addressed to Jin Kim, D'Amore-McKim School of Business, 360 Huntington Avenue, Boston, MA 02169, U.S.A. Email: jinkim@aya.yale.edu

**Abstract**

As large language models (LLMs) like ChatGPT become increasingly integrated into our everyday lives—from customer service and education to creative work and personal productivity—understanding how people interact with these AI systems has become a pressing issue. Despite the widespread use of LLMs, researchers lack standardized tools for systematically studying people's interactions with LLMs. To address this issue, we introduce *GPT for Researchers (G4R)*, or g4r.org, a free website that researchers can use to easily create and integrate a GPT Interface into their studies. At g4r.org, researchers can (1) enable their study participants to interact with GPT (such as ChatGPT), (2) customize GPT Interfaces to guide participants' interactions with GPT (e.g., set constraints on topics or adjust GPT's tone or response style), and (3) capture participants' interactions with GPT by downloading data on messages exchanged between participants and GPT. By facilitating study participants' interactions with GPT and providing detailed data on these interactions, G4R can support research on topics such as consumer interactions with AI agents or LLMs, AI-assisted decision-making, and linguistic patterns in human-AI communication. With this goal in mind, we provide a step-by-step guide to using G4R at g4r.org.

*Keywords:* large language model, LLM, GPT, ChatGPT, research methods, behavioral science, AI-assisted research

**How to Capture and Study Conversations Between Research Participants and ChatGPT:**

**GPT for Researchers (g4r.org)**

As large language models (LLMs) become increasingly popular and deeply embedded in both research and everyday life, behavioral scientists have shown growing interest in understanding how people interact with these models (e.g., ChatGPT). Despite this growing interest, there seems to be a lack of standardized tools for examining how people interact with LLMs. Many existing studies have relied on ad hoc solutions or custom-built platforms. For example, in an early study conducted shortly after ChatGPT's release, some researchers resorted to having participants sign up for ChatGPT in the middle of the study to use it in the subsequent stage of the study (Noy & Zhang, 2023), while other research teams built different platforms to investigate their respective research questions (Costello et al., 2024; Jelson et al., 2025; Nie et al., 2024; Wang et al., 2024). The diverse ways to study human-AI interactions may have made it difficult to compare findings across different research papers, replicate results from those papers, or efficiently scale research efforts. Against this backdrop, we introduce GPT for Researchers or G4R—a free and easy-to-use tool hosted at g4r.org—that enables researchers to (1) facilitate conversations between study participants and ChatGPT, (2) steer these conversations for their specific research aims, and (3) capture these conversations in their entirety.

Across various fields, different research teams have integrated LLMs into their studies— often developing custom tools tailored to their specific research objectives. For example, to study the impact of LLM use on coding education, Nie and colleagues conducted an experiment in an online coding class, for which they built a "course-specific interface to GPT-4" (Nie et al., 2024). Their interface featured "varying color schemes" to be "differentiated" from the original ChatGPT and presented course policy and sometimes example coding questions on the interface (Nie et al., 2024). To study how students use ChatGPT for writing essays, Jelson and colleagues built their own "writing-ChatGPT platform," a custom application featuring two tabs that simulated a web browser, one tab for writing essays and another for interacting with ChatGPT (Jelson et al., 2025). As yet another example, Costello and colleagues investigated how dialogues with an LLM can reduce conspiracy beliefs by writing JavaScript code (that called OpenAI's API) into a Qualtrics survey and by "dynamically inject[ing] participant-specific information into the [LLM's] instructions" (Costello et al., 2024). Finally, Wang and colleagues developed a "task platform that emulates the official ChatGPT interface" to study how supportive functions like prompt suggestions affect user experience (Wang et al., 2024). Their interface displayed things like follow-up questions and revised prompts with improved clarity, which were specifically designed for the researchers to examine how these supportive functions affected user experience with ChatGPT. All these research teams created their own versions of ChatGPT interface, specifically designed to answer their respective research questions.

Beyond those who successfully built custom tools, many researchers may have been discouraged from pursuing LLM-related research altogether due to the absence of a reliable, ready-to-use solution. Our informal discussions with colleagues at various institutions suggest that many researchers were interested in facilitating and examining participants' interactions with LLMs or AI systems but were discouraged from conducting such studies partly due to the lack of an easy and accessible solution. Moreover, discussions with various research groups reinforced our belief that developing a tool like G4R would be highly beneficial for researchers exploring various aspects of human-AI interactions. Recognizing this gap between researchers' needs and the lack of available tools, we set out to create g4r.org to provide a simple and accessible solution for facilitating and capturing participant-LLM interactions.

In the remainder of this paper, we discuss (1) what researchers can do when they use G4R for the first time (see the section "How to Use G4R); (2) how they can create and customize a GPT Interface for their study with a specific research goal in mind (see the section "How to Customize a GPT Interface"); (3) how they can integrate a GPT Interface into their studies (see the section "How to Integrate the GPT Interface in a Qualtrics Survey"); and (4) how they can capture and use the data on messages exchanged between participants and GPT (see the section "How to Download and Merge Message Data"); we then conclude.

## How to Use G4R

**Trying It Out**

To use G4R, a researcher first needs to visit g4r.org. The first page will show three main actions that they can take: "Create a GPT Interface," "Sign in," and "Create an account" (Figure 1). If they click on the top button, "Create a GPT Interface," they will be directed to a page for creating a new GPT Interface as a guest, which will be similar to the page shown in Figure 2. Here, the researcher can enter the desired features of the GPT Interface or simply accept all the default values and click the "Create this GPT Interface" button at the bottom of the page. Clicking this button will take them to a page confirming the details of the newly created GPT Interface. Below these details will be a "Preview the Interface" button, clicking which will open a new browser tab that shows a chat window titled "ChatGPT Interface for Prolific Studies" (see Figure 3, Panel A). The researcher can test out their customized GPT Interface by sending messages to GPT and receiving replies from it.

**Figure 1**

*The First Page of G4R (https://g4r.org)*

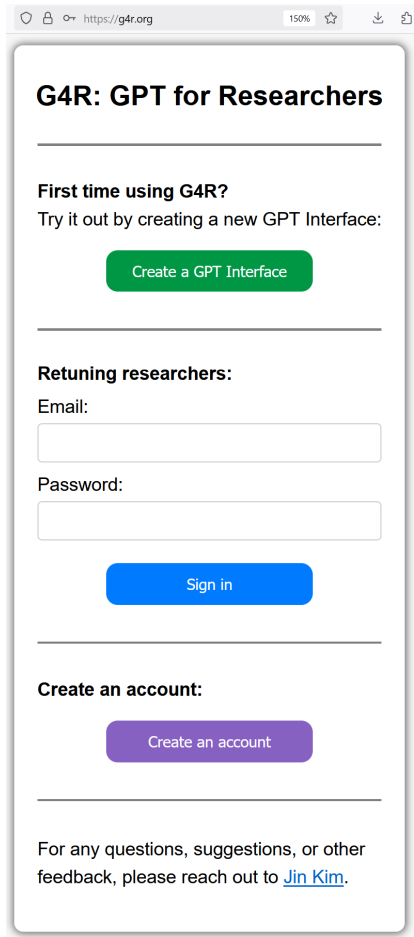

**Figure 2**

*Creating a GPT Interface on G4R*

**Creating an Account**

Once the researcher has experienced the procedure for creating a GPT Interface as a guest on g4r.org and has tested that the resulting GPT Interface functions as expected, we would recommend that they create a G4R account, which would allow them to create and customize multiple GPT Interfaces for themselves and download the message data for each of the interfaces they have created. To create an account, they can simply navigate back to the G4R main page (e.g., by clicking the "Go Back to G4R Home" button) and click the "Create an account" button at the bottom. Creating an account is straightforward: Researchers simply need to enter their name or nickname, email address, and password. As noted on the account creation page, the password they enter will be securely hashed according to the industry standard such that we as the operators of G4R cannot find out what password has been entered (G4R Operators & ChatGPT, 2025). After creating an account, the researcher can click on "Go to Your Researcher Home" button, which will take them to the "Researcher Home" page. This will be the same first page that a researcher will land on when they sign in from g4r.org. See the "Researcher Home" section below.

**Signing In**

The final action that a researcher can take on the main page of g4r.org is to sign into their account by entering their email address and password and clicking the button in the middle, "Sign In." Once they sign in, they will be directed to the "Researcher Home" page.

**Researcher Home**

"Researcher Home" is the main page where a researcher who signed in can take two primary actions: (1) creating (customizing) a GPT Interface and (2) downloading message data for each of the GPT Interfaces they have created. The first action they can take is creating a GPT

Interface. To create a GPT Interface, a researcher can follow the same procedure as they did when trying out G4R for the first time as a guest. That is, they can enter or update details of a GPT Interface or accept default values to customize a GPT Interface for themselves (see "How to Customize a GPT Interface" section below). The other action they can take on the "Researcher Home" page is downloading the message data for each of the GPT Interfaces they have created. In this second section of the page, a researcher will see a list of all the GPT Interfaces they have ever created on g4r.org along with a download link for each of the interfaces. They will be able to download the message data for any of their interfaces by clicking the links corresponding to the interfaces (see "How to Download Message Data" section below).

## How to Customize a GPT Interface

To customize a particular instance of a GPT Interface, a researcher can answer up to 12 questions either by accepting default values or entering new values. Entering a new answer to any of these questions, however, is optional, except for the first question (name of the study). The easiest thing for a researcher to do is to accept all the default values, which will create a GPT Interface that can be used in a typical study in which participants can interact with ChatGPT for simple tasks. However, we recommend that a researcher consider each of the 12 ways below for customizing their GPT Interface so that they can create a GPT Interface that is most suitable for their study design.

**Name of the Study**

The first question on the GPT Interface creation page asks the researcher, "What is the name of your study? (Participants will not see the name)." They can answer this question by entering a study name that will be associated with their new GPT Interface. They can enter any name they want (up to 300 characters long), and this will be the study name they will later be used to identify the particular instance of GPT Interface when they download message data for that GPT Interface. This name will not be visible to study participants. This question will not appear when a researcher tries out creating a GPT Interface as a guest (i.e., when they are not signed in).

**G4R Access Mode: New Browser Tab or Embedded Within a Qualtrics Survey**

The second question asks about one of the most important features of the new instance of their GPT Interface: "How do you want participants to access the GPT interface?" The researcher will choose how the GPT Interface will be accessed by their study participants—either on a new browser tab or embedded within a Qualtrics survey. If they choose the default, first option ("In a New Browser Tab"), the particular instance of GPT Interface will be integrated into a Qualtrics survey question as a prominent green button. If a study participant clicks the green button, then a new browser tab will open wherein the participant can interact with ChatGPT as shown in Figure 3, Panel A. On the other hand, if the researcher chooses the second option ("Embedded within Qualtrics"), then the particular instance of GPT Interface can be directly embedded within a Qualtrics survey question as shown in Figure 3, Panel B.

**Maximum Number of Messages**

The third question asks, "How many messages will each participant be able to send to GPT?" Participants can enter a number from 0 to 1,000 to set the maximum number of messages that each study participant will be able to send ChatGPT. When a study participant hits this maximum number of messages, they will see the message "You have sent the maximum allowed messages" and the button for sending a message will be disabled.

**Labels for Participants and GPT**

The fourth and fifth questions ask, "What is the label for the participant / GPT in the GPT Interface?" A researcher can enter a character string for each of these questions to customize the label that the study participant will see. For example, if the researcher accepts the default values ("You" and "ChatGPT"), the study participant will see "ChatGPT" as their counterpart in the GPT Interface and will see "You" as indicating the study participant themselves. Changing these labels would be suitable for certain study designs. For example, if a researcher wants to study customers' interaction with an AI customer representative, they could enter values like "Valued Customer" and "AI Customer Representative."

**System Prompt**

The sixth question is another one of the most important features as it allows the researcher to set the system prompt for the particular instance of GPT Interface. It asks, "(Optional) If you would like to send a system message to GPT to set guidelines and/or contexts for the conversation, enter it here ([What is this?](#))." A system prompt lets researchers guide ChatGPT's responses by setting things like the tone, context, or constraints for the study participants' interactions with GPT. It can shape interactions to fit study goals, such as encouraging creativity or simulating expert advice. If left blank, ChatGPT defaults to its general behavior, which may not suit studies that need more controlled interactions. Thoughtful use of this feature would ensure that the particular GPT Interface aligns with a specific research objective.

**First Message by GPT**

The seventh question asks, "(Optional) If you would like GPT to send the first message to initiate the conversation, enter it below." A researcher can enter any message here that will be sent as the first message from ChatGPT. The default message ("What can I help with?") is currently the first message shown to users of the original ChatGPT as of March 2025.

**Temperature for GPT**

The eight question asks, "(Optional) If you would like to adjust the temperature parameter, enter a value below." Temperature is a parameter that controls the randomness of ChatGPT's responses. Lower values (e.g., values closer to 0.0) make responses more focused and deterministic, while higher values (e.g., values closer to 2.0) make responses more random, diverse, or creative (OpenAI, 2025; ruv, 2023). The default value is 1.0, but researchers can enter any number between 0.0 and 2.0 (inclusive). If a researcher changes the temperature (i.e., enter any value other than 1.0), we strongly advise that the researcher tests out the GPT Interface multiple times to get a sense of randomness of GPT's responses before employing the particular instance of GPT Interface in their study.

**Text to Precede or Follow Each Message From Study Participants to GPT**

The ninth and tenth questions ask, "If you would like a text to precede / follow each message from participant to GPT, enter it here." Participants can enter any text for either of the two fields, and the entered text will precede or follow each of the messages that any study participant sends to GPT. For example, if the text to precede each message is "Please be concise," while the text to follow each message is "Thank you," then GPT will likely respond to each of any participant's message by trying to be concise and with the recognition that appreciation was expressed.

**OpenAI API Key**

The 11[th] question asks, "(Optional) If you want to use your own OpenAI API Key, please enter it here." A researcher can enter their own OpenAI API key for the particular instance of GPT Interface. Although entering an API key is currently optional (as of March 2025), researchers in the future may be asked to provide their own OpenAI API key if G4R becomes popular enough that the authors' API allowance nears its limit.

**HTML for the Top Section of the GPT Interface**

The 12$^{th}$ question asks, "(Optional) If you would like to customize the interface in the section above the conversation, enter the HTML for the section here." Researchers can enter HTML code in this section to customize the particular instances of GPT Interface. For example, if a researcher enters the HTML code ("<div style='background-color: blue; height: 20px'></div>"), then this code will result in a blue bar that is 20 pixels thick appearing at the top of the GPT Interface (i.e., in the section above the conversation). Such HTML code entered will be used only for GPT Interfaces that are accessed in a new browser tab, but not for the GPT Interfaces that are embedded wihin a Qualtrics survey.

**The Final Steps of Customizing a GPT Interface**

After answering or accepting default values for the questions above, the researcher can click on the "Create this GPT Interface" button at the bottom of the page. When they do so, they will next see a page that confirms all the details of the GPT Interface they have just customized. Below these details, they will see the green button labeled "Preview the Interface." When they click this button, a new browser tab will open and show a chat window with all the features of the customized GPT Interface.

## How to Integrate the GPT Interface in a Qualtrics Survey

In the last section of the page that confirms the details of the GPT Interface newly customized, researchers will see the heading "Next Steps." This section provides instructions for researchers on how to integrate the GPT Interface they just customized in their Qualtrics survey. These instructions are discussed and explained below.

**Copying the JavaScript Code**

The first step of integrating the customized GPT Interface involves copying and pasting the JavaScript code provided by G4R. Right below the first instruction ("Step 1"), a researcher will see a section of JavaScript code in gray background. To copy this code, the researcher can click the green button at the top right labeled "Click to copy this code." Clicking this button will copy the code into the clipboard of the researcher's operating system. Next, the researcher can navigate to the question in their Qualtrics survey where the GPT Interface will be integrated. For the given Qualtrics question, the researcher can click the "JavaScript" button within the "Edit Question" tab for the question. Doing so will open the JavaScript code window for the question with the default JavaScript code. The researcher can delete the entire default code and then paste the JavaScript code they had copied from g4r.org. The entire procedure of copying the JavaScript code is quite simple and is illustrated in a gif image on the G4R page, right below the JavaScript code, under the heading "How to copy the code." A researcher can also click the link that follows the heading ("see also the video") to watch a YouTube video that demonstrates how a researcher can copy the JavaScript code to integrate their newly created GPT Interface.

**Adding an Embedded Data Field to Store Participant IDs**

After copying the Javarscript code from G4R into the appropriate Qualtrics survey question, the researcher should add an embedded data field in the Qualtrics survey. They can do so by opening the "Survey Flow" for the Qualtrics survey, clicking on "Add a New Element Here" link at the bottom of the survey flow, clicking on "Embedded Data," and entering "g4r_pid" as the name of the new field. They should not set any value for this field such that the note from Qulatrics ("Value will be set from Panel or URL.") remains visible. The researcher should then move this embedded data field to the top of the Survey Flow, either by dragging and dropping the Embedded Data block at the top or moving it by pressing the "up" arrow key on the keyboard multiple times. Putting this Embedded Data field at the top of the Survey Flow ensures that this field will be initialized before any value will later be assigned to it. For each participant, this initially empty Embedded Data field will later store the participant ID that is randomly generated via the JavaScript code from Step 1 above. Importantly, this participant ID is what a researcher can later use to merge message data (which they will download from the "Researcher Home" page of g4r.org) with the survey response data (which they will later download from Qualtrics).

**Recruiting Participants and Collecting Data**

After a researcher follows the two steps above (i.e., copying the JavaScript code from g4r.org into a question in their Qualtrics survey and adding the embedded data field in the Survey Flow), they can recruit their study participants and collect data for their study, during which (some or all of) the study participants will interact with ChatGPT.

## How to Download and Merge Message Data

**How to Download Message Data**

The "Next Steps" section following the confirmation of the details of the newly created GPT Interface also provides the instructions on how to download and merge message data. First, to download the message data, a researcher signed into g4r.org can navigate to "Researcher Home" (i.e., the main page the researcher lands on after signing into G4R). In the lower section of the page, they will see the heading "Download G4R Messages Data." As instructed there, a researcher can view the table containing study names and download links, find their studies (i.e.,

GPT Interfaces) of interest, and click on the "Download" link, which will start downloading a CSV file of the message data. This message data set will contain each study participant's participant ID, each of the messages that each study participant sent to GPT using the given interface, each message they received from GPT, and the timestamp for each exchange in each study participant's conversation with GPT.

**How to Merge Message Data**

Researchers can merge (1) the message data downloaded from G4R with (2) the survey response data from Qualtrics using the participant IDs from the message data (i.e., 1) to match them with the values under "g4r_pid" column of the Qualtrics survey response data (i.e., 2). We can illustrate a tedious and discouraged method of merging these two datasets. Suppose that a researcher sees the first participant in their Qualtrics survey response data with the value "ABC" in the "g4r_pid" column. The researcher can open the message data and find all the rows whose Participant ID value is "ABC." If there were two messages that this study participant sent to GPT and two messages that they received from GPT, the researcher *could* create four new columns in the Qualtrics data as follows: "message_to_gpt_1," "message_from_gpt_1," "message_to_gpt_2," "message_from_gpt_2." Next, for each cell across these four columns, the researcher *could* copy and paste the appropriate messages. Of course, we discourage such a tedious method of merging thousands of message data across hundreds of participants.

Instead, to simplify the process of merging message data with Qualtrics response data, we provided three sample files that researchers can modify and use on the "Researcher Home" page, right above the table that displays study names and message data download links: (1) Sample message data, (2) Sample Qualtrics data, and (3) Sample R script for merging the two data sets. All that a researcher needs to do is download these three files, place them in the same folder, and execute the sample R script. As this R script is annotated, self-explanatory, and requires minimal use of additional packages (except for the two R packages, 'data.table' and 'kim' [Dowle & Srinivasan, 2021; Kim, 2024]), we expect that researchers would be able to easily update this R script for their particular study. For example, they can place the message data and Qualtrics survey data in the same folder as the sample R script and then open the sample R script. They can then change the file names for the message data and Qualtrics survey data in the sample R script, and execute the entire R script, which should produce a new data set in which the message data and Qulatrics survey data will be merged.

**Figure 3**

*A GPT Interface in a New Browser Tab (A) and Embedded Within Qualtrics Survey (B)*

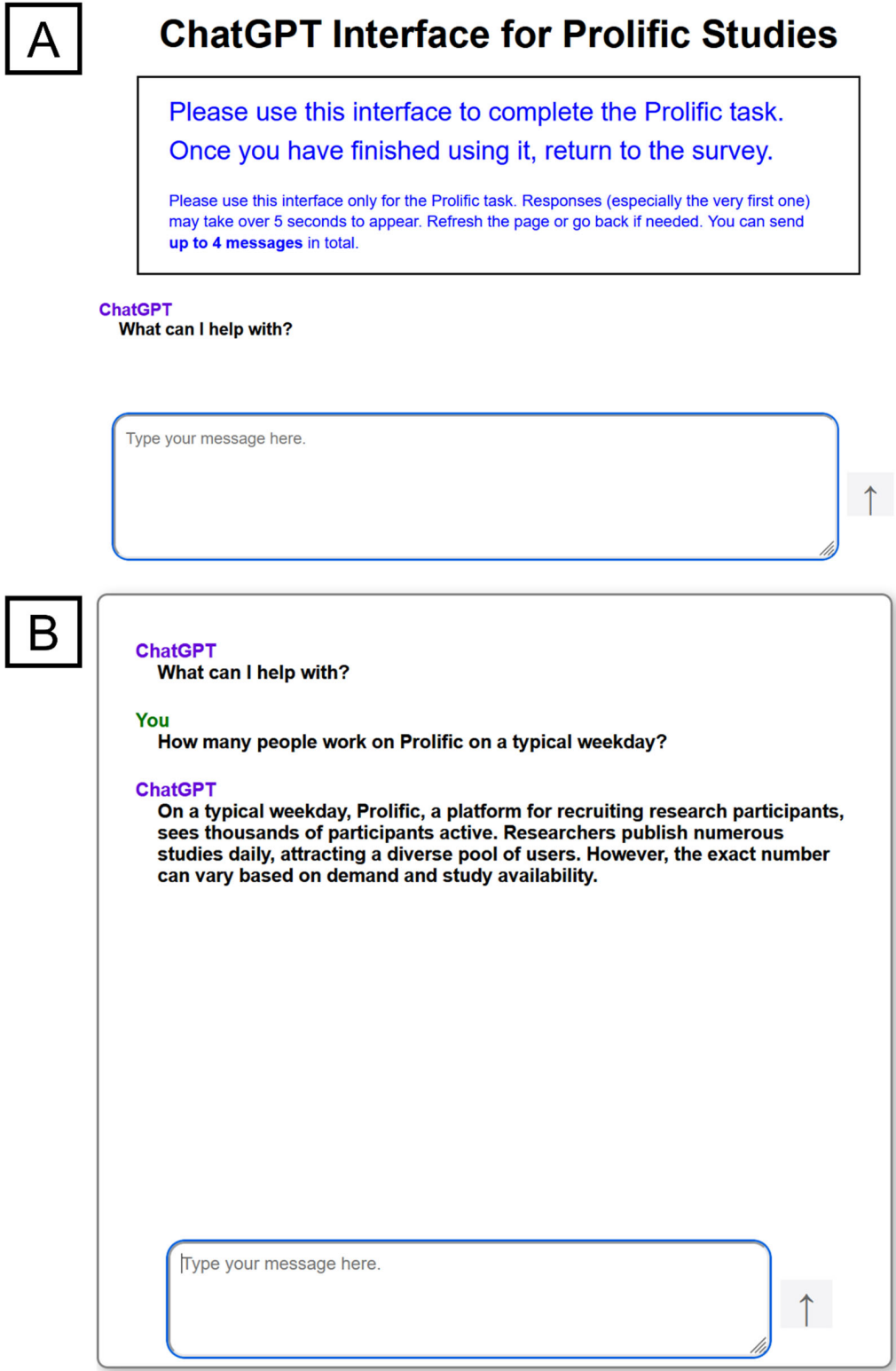

## Conclusion

We introduce GPT for Researchers (G4R) or g4r.org, a free and simple website that researchers can use to enable and capture study participants' interactions with ChatGPT in research studies. We demonstrate how researchers can use g4r.org to integrate a GPT Interface in their research studies, specifically by embedding it in their Qualtrics survey or by allowing study participants to open the interface in a new browser tab. More importantly, we show how a researcher can use g4r.org to customize a GPT Interface to fit their research goals. For example, a researcher can set labels for study participants and the AI model engaging in a conversation (e.g., "Valued Customer" and "AI Customer Representative") or customize the system prompt for their GPT Interface to guide the way that ChatGPT will interact with study participants (e.g., by setting the tone, context, or constraints for the interactions with GPT). In addition, we show how researchers can download the message data from g4r.orog and merge them with the survey response data from Qualtrics for further analyses on study participants' interactions with GPT. In sum, G4R (g4r.org) addresses the lack of standardized tools for studying human-AI interactions by allowing researchers to enable, customize, and capture study participants' interactions with GPT and by providing an easy-to-use and accessible solution for behavioral research focused on human-AI interactions. We encourage researchers to try out g4r.org and share their feedback (e.g., by contacting the corresponding author), so we can continue improving the tool and make future versions even more useful for behavioral research.